\documentclass[11pt]{article}

\usepackage{cite}    
\usepackage{graphicx}
\usepackage{subfigure} 
\usepackage{url}      
 \newcommand{\eqn}[1]{(\ref{#1})}
 \newcommand{\eqs}[1]{Eqs.~(\ref{#1})}
 \newcommand{\Fig}[1]{Figure~\ref{#1}}
 
 \newcommand{\fig}[1]{Fig.~\ref{#1}}

\begin{document}
 \title{Performance of DPSK Signals with Quadratic Phase Noise}

 \author{Keang-Po Ho%
 \thanks{This research was supported in part by the National Science Council of R.O.C. under Grant NSC-92-2218-E-002-034.
}
 \thanks{K.-P. Ho is with the Institute of Communication Engineering and Department of Electrical Engineering, 
 National Taiwan University, Taipei 106, Taiwan.
 (Tel: +886-2-2363-5251 ext. 222, Fax: +886-2-2368-3824, E-mail: kpho@cc.ee.ntu.edu.tw)}
 }

 \markboth{IEEE Transactions of Communications}{K.-P. Ho:Performance of DPSK Signals with Quadratic Phase Noise}

 \maketitle

 \begin{abstract}
 Nonlinear phase noise induced by the interaction of fiber Kerr effect and amplifier noises is a quadratic function of the electric field. 
When the dependence between the additive Gaussian noise and the quadratic phase noise is taking into account, the error probability for differential phase-shift keying (DPSK) signals is derived analytically.
Depending on the number of fiber spans, the signal-to-noise ratio (SNR) penalty is increased by up to 0.23 dB due to the dependence between the Gaussian noise and the quadratic phase noise.  \\
 \end{abstract}


 \section{Introduction}

Other than the projection of additive Gaussian noise to the phase, phase noises from other sources can be considered as multiplicative noise that adds directly to the phase of the received signal.
When the local oscillator is not locked perfectly into the signal, the noisy reference gives additive phase noise \cite{lindsey73, jain74}.
Laser phase noise degrades coherent optical communication systems \cite{nicholson84, foschini88, smith95}.
Those types of extra additive phase noise that add directly into the signal phase are independent of the additive Gaussian noise.
In this paper, the additive phase noise is quadratic function of the electric field.
When the electric field is contaminated with additive Gaussian noise, although the quadratic phase noise is uncorrelated with the linear phase noise, both non-Gaussian distributed, the phase noise weakly depends on the additive Gaussian noise.

Differential phase-shift keying (DPSK) signals \cite{gnauck02, gnauck03, rasmussen03, zhu03, cho03, xu03, kim03a, wree03a, mizuochi03, cai04, gnauck04}  have received renewed attention recently for long-haul or spectrally efficiency lightwave transmission systems.
 When optical amplifiers are used periodically to compensate the fiber loss, the interaction of optical amplifier noise and fiber Kerr effect induced nonlinear phase noise, often called Gordon-Mollenauer effect \cite{gordon90}, or more precisely, nonlinear phase noise induced by self-phase modulation.
 Added directly into the signal phase, Gordon-Mollenauer effect is a quadratic function of the electric field and degrades DPSK signal \cite{gordon90, ryu92, mecozzi94, xu03, kim03, ho03sta, ho0309a, ho0309b, mizuochi03}.

Previous studies found the variance or the corresponding $Q$-factor of the quadratic phase noise \cite{gordon90, xu02, liu02, xu03, ho0306, ho0403a} or the spectral broadening of the signal \cite{ryu92, saito93, mizuochi03}. 
Recently, quadratic phase noise is found to be non-Gaussian distributed both experimentally \cite{kim03} and theoretically \cite{ho0308, ho0309c}. 
As non-Gaussian random variable, neither the variance nor $Q$-factor is sufficient to completely characterize the phase noise.
The probability density of quadratic phase noise is found in \cite{ho0309c} and used in \cite{ho0309b} to evaluate the error probability of DPSK signal by assuming that quadratic phase noise and Gaussian noise are independent of each other.
However, as shown in the simulation of \cite{ho0309b, ho0309a}, the dependence between Gaussian noise with quadratic phase noise increases the error probability.

Using the distributed assumption of infinite number of fiber spans, the joint statistics of nonlinear phase noise and Gaussian noise is derived analytically by \cite{mecozzi94, mecozzi94a, ho03sta}.
The characteristic function of nonlinear phase noise becomes a very simple expression with the distributed assumption \cite{ho0308}.
The error probability of DPSK signal has been derived with \cite{ho0309a} and without \cite{ho03sta, ho03exa} the assumption that nonlinear phase noise is independent of the Gaussian noise.
Based on the distributed assumption, it is found that the dependence between linear and nonlinear phase noise increases both the error probability and SNR penalty \cite{ho03sta, ho03exa}. 

The distributed assumption is very accurate when the number of fiber spans is larger than 32 \cite{ho0308, ho03sta}.
For a typical fiber span length of 80 km, a fiber link of 32 spans has a total distance of over 2500 km.
Most terrestrial fiber systems have an overall distance of less than 1000 km, the distributed assumption needs to be verified for small number of fiber spans.
Recently, DPSK signals have been used in systems with small number of fiber spans \cite{miyamoto02, bissessur03, gnauck04}. 
Of course, the independence assumption can be used for either small \cite{ho0309b} or large \cite{ho0309a} number of fiber spans.
However, as shown in \cite{ho03sta, ho03exa}, the independence assumption of \cite{ho0309a, ho0309b} underestimates both the error probability and the required SNR, contradicting to the principles of conservative system design.

In this paper, taking into account the dependence between the quadratic phase noise and Gaussian noise, the error probability of DPSK signal is derived for finite number of fiber spans, to our knowledge, the first time.
Comparing with the independence approximation of \cite{ho0309b}, the dependence between the quadratic phase noise and Gaussian noise increases the error probability of the system.

In the remaining parts of this paper, Sec. \ref{sec:quad} gives the model of the quadratic phase noise, mostly follows the approaches of \cite{ho0309c}; Sec. \ref{sec:joint} derives the joint statistics of the additive Gaussian noise and the quadratic phase noise; Using the joint statistics, Sec. \ref{sec:ber} gives the exact error probability of DPSK signals with quadratic phase noise, taking into account the dependence between the additive Gaussian noise and quadratic phase noise; Sec. \ref{sec:num} calculates the error probability and the SNR penalty of DPSK signals, and compared with the independence approximation of \cite{ho0309b}; Sec. \ref{sec:end} is the conclusion of the paper. 

\section{Quadratic Nonlinear Phase Noise}
\label{sec:quad}
 
For an $N$-span systems, for simplicity and without loss of generality, the overall quadratic phase noise is \cite{ho0309c, gordon90, liu02, ho0306}

 \begin{eqnarray}
 \Phi_{\mathrm{NL}} & = & |\vec{E}_0 + \vec{n}_1|^2 +
	  |\vec{E}_0 +\vec{n}_1 + \vec{n}_2|^2  \nonumber \\
        & & \qquad \qquad + \cdots + |\vec{E}_0 + \vec{n}_1 + \cdots + \vec{n}_N|^2,
 \label{phiNL}
 \end{eqnarray}

\noindent where $\vec{E}_0 = (A, 0)$ is a two-dimensional vector as the baseband representation of the transmitted electric field, $\vec{n}_k, k = 1, \dots, N$, are independent identically distributed (i.i.d.) zero-mean circular Gaussian random complex numbers as the optical amplifier noise introduced into the system at the $k^{\mathrm{th}}$ fiber span.
 Both electric field of $\vec{E}_0$ and amplifier noises of $\vec{n}_k$ in \eqn{phiNL} can also be represented as complex number.
The variance of $\vec{n}_k$ is $E\{|\vec{n}_k|^2\} = 2 \sigma_0^2$, $k = 1, \dots, N$, where $\sigma_0^2$ is the noise variance per span per dimension. 
In \eqn{phiNL}, the constant factor of the product of fiber nonlinear coefficient and the effective nonlinear length per span, $\gamma L_\mathrm{eff}$, is ignored for simplicity. 
Without affected the SNR, both signal and noise in \eqn{phiNL} can be scaled by the same ratio for different mean nonlinear phase shift of $<\!\!\Phi_{\mathrm{NL}}\!\!> = NA^2 + N (N+1) \sigma_0^2$ except the case without quadratic phase noise of $<\!\!\Phi_{\mathrm{NL}}\!\!> = 0$.
After the scaling, the mean nonlinear phase shift is approximately equal to the product of number of fiber spans and the launched power per span, especially for the usual case of large SNR with small noise.

In the linear regime, ignoring the fiber loss of the last span and the amplifier gain required to compensate it, the signal received after $N$ spans is 

\begin{equation}
\vec{E}_N = \vec{E}_0 + \vec{n}_1 + \vec{n}_2 + \cdots + \vec{n}_N
\label{EN}
\end{equation}
 
\noindent with a power of $P_N = |\vec{E}_N|^2$ and SNR of $\rho_s = A^2/(2 N \sigma_0^2)$. 
In \eqs{phiNL} and (\ref{EN}), the configuration of each fiber spans is assumed to be identical with the same length and launched power.

In \cite{ho0309c}, using the method of \cite{kac47, turin60}, the characteristic function of the quadratic phase noise \eqn{phiNL} is found to be

\begin{equation}
 \Psi_{\Phi_{\mathrm{NL}}}(\nu)  = \prod_{k = 1}^N \frac{1}{1 - 2 j \nu \sigma_0^2 \lambda_k} 
	\exp \left[ \frac {j \nu A^2 (\vec{v}_k^T \vec{w})^2/\lambda_k} {1 - 2 j \nu \sigma_0^2 \lambda_k} \right].
\label{cfPhiNL} 
\end{equation}

\noindent where $\vec{w} = (N, N-1, \ldots, 2, 1)^T$,  $\lambda_k$, $\vec{v}_k$, $k = 1, 2, \ldots, N$ are the eigenvalues and eigenvectors of the covariance matrix ${\mathcal C }$, respectively.
The covariance matrix ${\mathcal C} = {\mathcal M}^T {\mathcal M}$ with

\begin{equation}
{\mathcal M} = \left[ 
\begin{array}{ccccc}
1 & 0 & 0 & \cdots & 0 \\
1 & 1 & 0 & \cdots & 0 \\
1 & 1 & 1 & \cdots & 0 \\
\vdots & \vdots & \vdots & \ddots & \vdots \\
1 & 1 & 1 & \cdots & 1
\end{array} 
\right].
\end{equation}

The characteristic function of \eqn{cfPhiNL} is used to find the error probability of a DPSK signal in \cite{ho0309b} based on the assumption that the quadratic phase noise of \eqn{phiNL} is independent of the received electric field of \eqn{EN}.

\section{Joint Statistics of Gaussian Noise and Quadratic Phase Noise}
\label{sec:joint}

To find the dependence between the quadratic phase noise and the received electric field, the joint characteristic function of

\begin{equation}
\Psi_{\Phi_{\mathrm{NL}}, \vec{E}_N}(\nu, \vec{\omega}) = 
    E\left\{ \exp( j \nu \Phi_{\mathrm{NL}} + j \vec{\omega} \cdot \vec{E}_N \right\}
\end{equation}

\noindent will be derived here with $\Phi_{\mathrm{NL}}$ and $\vec{E}_N$ given by \eqn{phiNL} and \eqn{EN}, respectively.

Similar to \cite{ho0309c, ho03sta}, with $\vec{\omega} = (\omega_1, \omega_2)$ and $\vec{E}_N = (e_1, e_2)$, we obtain

\begin{eqnarray}
&  j \nu \varphi_1 + j \omega_1 e_1 \nonumber \\
&  = j \nu NA^2 + j \omega_1 A + 2j \nu A \vec{w}^T \vec{x} + j \omega_1 \vec{w}^T_I \vec{x}  +j \nu \vec{x}^T {\mathcal C} \vec{x}, \nonumber \\
\label{varphi1e1Mat}
\end{eqnarray}

\noindent where $\varphi_1$ is given by

\begin{equation}
\varphi_1 = |A + x_1|^2 + |A + x_1 + x_2|^2 + \cdots + |A + x_1 + \cdots + x_N|^2,
\label{varphi1}
\end{equation}

\noindent with $\vec{n}_i = (x_i, y_i)$, $i = 1, \dots, N$,
$\vec{w}_I = (1, 1, \ldots, 1)^T$,  

\[
j \omega_1 e_1 = j \omega_1 (A + x_1 + x_2 + \cdots + x_N) = j \omega_1 A + j \omega_1 \vec{w}_I^T \vec{x},
\]

\noindent and $\vec{x} = (x_1, x_2, \dots, x_N)^T$.

Similar to \cite{ho0309c}, using the $N$-dimensional Gaussian probability density function (p.d.f.)~of ${(2 \pi \sigma_0^2)^{-\frac{N}{2}}}\exp \left( - { \vec{x}^T \vec{x}}/{2 \sigma_0^2} \right)$ for $\vec{x}$, we obtain

\begin{eqnarray}
\lefteqn{ \Psi_{\varphi_1, e_1}(\nu, \omega_1)
   =  \frac{e^{j \nu N A^2 + j \omega_1 A } } {(2 \pi \sigma_0^2)^{\frac{N}{2}}} } \nonumber \\
 & & \times			\int \exp \left[ 2 j \nu A \vec{w}^T \vec{x} +  j \omega_1 \vec{w}^T_I \vec{x} -
		\vec{x}^T \Gamma \vec{x} \right] \mathrm{d} \vec{x}.
\label{cfvarphi1e1}
\end{eqnarray}

\noindent or 

\begin{eqnarray}
\lefteqn{ \Psi_{\varphi_1, e_1}(\nu, \omega_1)\!
  = \! e^{j \nu N A^2 + j \omega_1 A }  {(2 \sigma_0^2)^{-\frac{N}{2}}} { \det[\Gamma]^{-\frac{1}{2}}} } \nonumber \\
& & \!\! \times \exp{\left[ -\left(\nu A \vec{w}+ \frac{1}{2}\omega_1 \vec{w}_I\right)^T \Gamma^{-1} \left(\nu A \vec{w}+ \frac{1}{2}\omega_1 \vec{w}_I\right) \right]}, \nonumber \\
\label{cfVarphi1e1} 
\end{eqnarray}

\noindent where  $\Gamma = {{\mathcal I}/(2 \sigma_0^2) - j \nu  {\mathcal C }}$ and ${\mathcal I}$ is an $N \times N$ identity matrix. 

\noindent Similarly, using $A = 0$ in \eqn{cfVarphi1e1}, we get

\begin{equation}
\Psi_{\varphi_2, e_2}(\nu, \omega_2)
  = \frac{\exp{\left[ - \frac{1}{4} \omega^2_2 \vec{w}_I^T \Gamma^{-1}\vec{w}_I \right]}}
  { {(2 \sigma_0^2)^{\frac{N}{2}}} { \det[\Gamma]^{\frac{1}{2}}}}. 
\end{equation}

\noindent where 

\begin{equation}
\varphi_2 = y_1^2 + |y_1 + y_2|^2 + \cdots + |y_1 + \cdots + y_N|^2.
\label{varphi2}
\end{equation}

The joint characteristic function of 

\begin{equation}
\Psi_{\Phi_{\mathrm{NL}}, \vec{E}_N}(\nu, \vec{\omega})
   = \Psi_{\varphi_1, e_1}(\nu, \omega_1) \Psi_{\varphi_2, e_2}(\nu, \omega_2)
\end{equation}

\noindent becomes

\begin{equation}
\Psi_{\Phi_{\mathrm{NL}}, \vec{E}_N}(\nu, \vec{\omega})
   = \Psi_{\Phi_\mathrm{NL}}(\nu) \exp\left[ j \omega_1 m_N(\nu) - \sigma_N^2(\nu) \frac{|\vec{\omega}|^2}{2} \right],
\label{cfPhiNLEN}
\end{equation}

\noindent where

\begin{eqnarray}
\Psi_{\Phi_\mathrm{NL}}(\nu) &=& \frac{\exp{\left[ j \nu N A^2 -\nu^2 A^2 \vec{w}^T \Gamma^{-1} \vec{w} \right]}}
  {(2 \sigma_0^2)^{N}\det[\Gamma]}, \\
m_N(\nu) &=& A + j \nu A  \vec{w}^T \Gamma^{-1} \vec{w}_I, \\
\sigma_N^2(\nu) & = & \frac{1}{2} \vec{w}_I^T \Gamma^{-1} \vec{w}_I.
\end{eqnarray}

Based on the eigenvalues and eigenvectors of the covariance matrix ${\mathcal C }$, the characteristic function of $\Psi_{\Phi_\mathrm{NL}}(\nu)$ becomes that of \eqn{cfPhiNL}, and 

\begin{eqnarray}
m_N(\nu) & = & A  + 2 j \nu \sigma_0^2 A \sum_{k=1}^N \frac{ (\vec{v}_k^T \vec{w})(\vec{v}_k^T \vec{w}_I)}
                                         {1 - 2 j \nu \sigma_0^2 \lambda_k} \nonumber \\
& = & A \sum_{k=1}^N \frac{ (\vec{v}_k^T \vec{w})(\vec{v}_k^T \vec{w}_I)/\lambda_k}
                                         {1 - 2 j \nu \sigma_0^2 \lambda_k},
\label{meanNnu} \\
\sigma_N^2(\nu) & = & \sigma_0^2 
                   \sum_{k = 1}^N \frac{ (\vec{v}_k^T \vec{w}_I)^2}
                                         {1 - 2 j \nu \sigma_0^2 \lambda_k}.
\label{sigmaNnu}
\end{eqnarray}

The characteristic function of \eqn{cfPhiNLEN} is similar to the corresponding characteristic function with the distributed assumption \cite{ho03sta}. 
If the number of spans $N$ approaches infinite, the characteristic function should converge to that of \cite{ho03sta}.

Based on \eqn{cfPhiNLEN}, we obtain

\begin{equation}
\mathcal{F}^{-1}_{\vec{\omega}} \left\{ \Psi_{\Phi_{\mathrm{NL}}, \vec{E}_N} \right\}   =   \frac{\Psi_{\Phi_\mathrm{NL}}(\nu)}{ 2 \pi \sigma_N^2(\nu)} \exp\left( -\frac{|\vec{z} -  \vec{\xi}_\nu|^2}{2 \sigma_N^2(\nu)} \right), 
\label{cfPhiNLpdfEN}
\end{equation}

\noindent with $\vec{\xi}_\nu= (m_N(\nu), 0)$, and $\mathcal{F}^{-1}_{\vec{\omega}}\{ \cdot \}$ denotes inverse Fourier transform with respect to $\vec{\omega}$.
The partial characteristic function and p.d.f. of \eqn{cfPhiNLpdfEN} is similar to a two-dimensional Gaussian p.d.f. with mean of $(m_N(\nu), 0)$ and variance of $\sigma^2_N(\nu)$.
With the dependence on the quadratic phase noise, the variance of $\sigma_N^2(\nu)$ and the mean of $m_N(\nu)$ are both complex numbers depending on the ``angular frequency'' of $\nu$. 
The marginal p.d.f. of the received electric field $\vec{E}_N$ is a two-dimensional Gaussian distribution with variance of $\sigma_N^2(\nu)|_{\nu = 0} = N \sigma_0^2$ and mean of $m_N(\nu)|_{\nu = 0} = A$.

With normalization, the corresponding joint characteristic of \eqn{cfPhiNLpdfEN} in \cite{ho03sta} has 

\begin{equation}
\sigma_{\infty}^2(\nu) = \frac{1}{2} \frac{\tan( \sqrt{j \nu})}{\sqrt{j \nu}} \mbox{ and } m_{\infty}(\nu) = \sec(\sqrt{j \nu}) \sqrt{\rho_s}
\end{equation}

\noindent when $N \rightarrow \infty$.
Based on joint statistics of \eqn{cfPhiNLpdfEN}, similar to that of \cite{ho03sta, ho03exa, ho03lc}, the exact error probability of DPSK signal can be derived analytically, even for case with linearly compensated nonlinear phase noise \cite{liu02, xu02, xu02a, ho0403a, ho0309b}.
As shown in \cite{ho03sta}, the optimal compensation curve of \cite{ho0403a, ho0306} can also be derived using \eqn{cfPhiNLpdfEN}.

\section{Exact Error Probability}
\label{sec:ber}

With nonlinear phase noise, assuming zero transmitted phase, the overall received phase is

\begin{equation}
\Phi_r = \Theta_n - \Phi_{\mathrm{NL}} 
\label{PhiRNL}
\end{equation}

\noindent where $\Theta_n$ is the phase of $E_N$ \eqn{EN}. 
The received phase is confined to the range of $[-\pi, +\pi)$. 
The p.d.f. of the received phase is a periodic function with a period of $2 \pi$.
If the characteristic function of the received phase is $\Psi_{\Phi_r}(\nu)$, the p.d.f. of the received phase has a Fourier series expansion of

\begin{equation}
p_{\Phi_r}(\theta) = \frac{1}{2 \pi} + \frac{1}{\pi} 
\sum_{m =1}^{+\infty}\Re \left\{ \Psi_{\Phi_r}(m) \exp( -j m \theta) \right\},
\label{pdfPhiR}
\end{equation}

\noindent where $\Re\{\cdot\}$ denotes the real part of a complex number.
In \eqn{pdfPhiR}, we use the conjugate symmetry property of  $\Psi_{\Phi_r}(-\nu) = \Psi^*_{\Phi_r}(\nu)$.

In order to derive the Fourier coefficient of $\Psi_{\Phi_r}(m)$, we need the joint characteristic function of $\Theta_n$ and $\Phi_{\mathrm{NL}}$ at integer ``angular frequency'' of $\nu = m$.
Based on \eqn{cfPhiNLpdfEN}, using the same method as \cite{prabhu69, mecozzi94, ho03sta, ho03exa}, we obtain

\begin{eqnarray}
\lefteqn{\Psi_{\Phi_{\mathrm{NL}}, \Theta_n}(\nu, m)   = \frac{\sqrt{\pi}}{2}\Psi_{\Phi_\mathrm{NL}}(\nu) \sqrt{\gamma(\nu)} e^{-\gamma(\nu) /2} } \nonumber \\
& &  \qquad \qquad \times \left\{ I_{\frac{m-1}{2}} \left[\frac{\gamma(\nu)}{2} \right] 
			+ I_{\frac{m-1}{2}} \left[\frac{\gamma(\nu)}{2} \right] \right\},  \nonumber \\
& & \qquad \qquad \qquad \qquad m \geq 0, 
\label{cfPhiThetam}
\end{eqnarray}

\noindent where $\gamma(\nu) = \frac{1}{2} m_N(\nu)^2/\sigma_N^2(\nu)$ is the complex-valued frequency dependence SNR parameter.
When $\nu = 0$, it is obvious that $\gamma(\nu)|_{\nu = 0} = \rho_s$.

From \eqn{PhiRNL}, the Fourier coefficient in \eqn{pdfPhiR} is $\Psi_{\Phi_r}(m) = \Psi_{\Phi_{\mathrm{NL}}, \Theta_n}(m, m)$.
For DPSK signal, the differential received phase is $\Delta \Phi_r = \Phi_r(t) -\Phi_r(t - T)$ in which the p.d.f.'s of $\Phi_r(t)$ and $\Phi_r(t - T)$ are the same as that of \eqn{pdfPhiR}.
The p.d.f. of the differential received phase is the same as \eqn{pdfPhiR} with Fourier coefficient equal to $|\Psi_{\Phi_r}(m)|^2$, i.e.,

\begin{equation}
p_{\Delta \Phi_r}(\theta) = \frac{1}{2 \pi} + \frac{1}{\pi} 
\sum_{m =1}^{+\infty}|\Psi_{\Phi_r}(m)|^2 \cos( m \theta).
\label{pdfDeltaPhiR}
\end{equation}

Similar to the procedure of \cite{jain73, jain74, blachman81, nicholson84, ho0309b, ho03sta, ho03exa, prabhu69}, the error probability becomes

\begin{equation}
p_e = \frac{1}{2} - \frac{2}{\pi}
\sum _{k =0}^{+\infty}
   \frac{(-1)^k}{2 k + 1} \left| \Psi_{\Phi_r}(2 k + 1) \right|^2.
\end{equation}

\noindent or

\begin{eqnarray}
\lefteqn{p_e =\frac{1}{2} - \frac{1}{2} 
	  \sum_{k = 0}^{\infty} \frac{(-1)^k \left| r_k e^{-r_k} \right|} {2 k + 1}
    \left| I_{k}\!\!\left(\frac{r_k}{2} \right) 
			+ I_{k+1}\!\!\left(\frac{r_k}{2} \right) \right|^2 }\nonumber \\
& &  \qquad \qquad \qquad \qquad \qquad \quad \times \left| \Psi_{\Phi_{\mathrm{NL}}}(2 k + 1) \right|^2, \qquad 
\label{BERDPSK}
\end{eqnarray}

\noindent where

\begin{equation}
r_k = \frac{m^2_N(2k+1)}{2 \sigma^2_N(2k+1)}
\label{lambdak}
\end{equation}

\noindent analogous to the ``angular frequency'' depending SNR as the ratio of complex power of $\frac{1}{2}m^2_N(\nu)$ to the noise variance of $\sigma^2_N(\nu)$.

The error probability expression of \eqn{BERDPSK} is almost the same as that in \cite{ho03exa, ho03sta} but with a different parameter of \eqn{lambdak}.
The error probability of \eqn{BERDPSK} is also similar to the cases when additive phase noise is independent to Gaussian noise \cite{jain73, jain74, blachman81, nicholson84, ho0309b}. 
The frequency depending SNR is originated from the dependence between the additional phase noise and the Gaussian noise \cite{mecozzi94, ho03sta, ho03exa, ho03lc}.

\section{Numerical Results}
\label{sec:num}

\begin{figure}
\centerline{\includegraphics[width = 0.5 \textwidth]{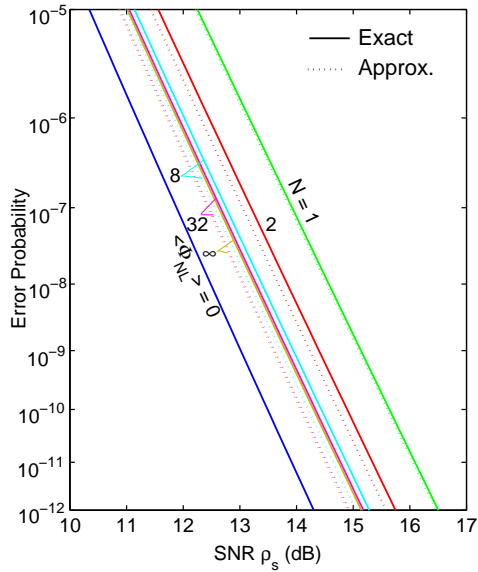}}
\caption{The error probability of DPSK signal as a function of SNR for $N = $1, 2, 4, 8, 32, and infinite number of fiber spans and mean nonlinear phase shift of $<\!\!\Phi_\mathrm{NL}\!\!> = 0.5$ rad.
}
\label{BerDPSKQ}
\end{figure}

\begin{figure}
\centerline{\includegraphics[width = 0.65 \textwidth]{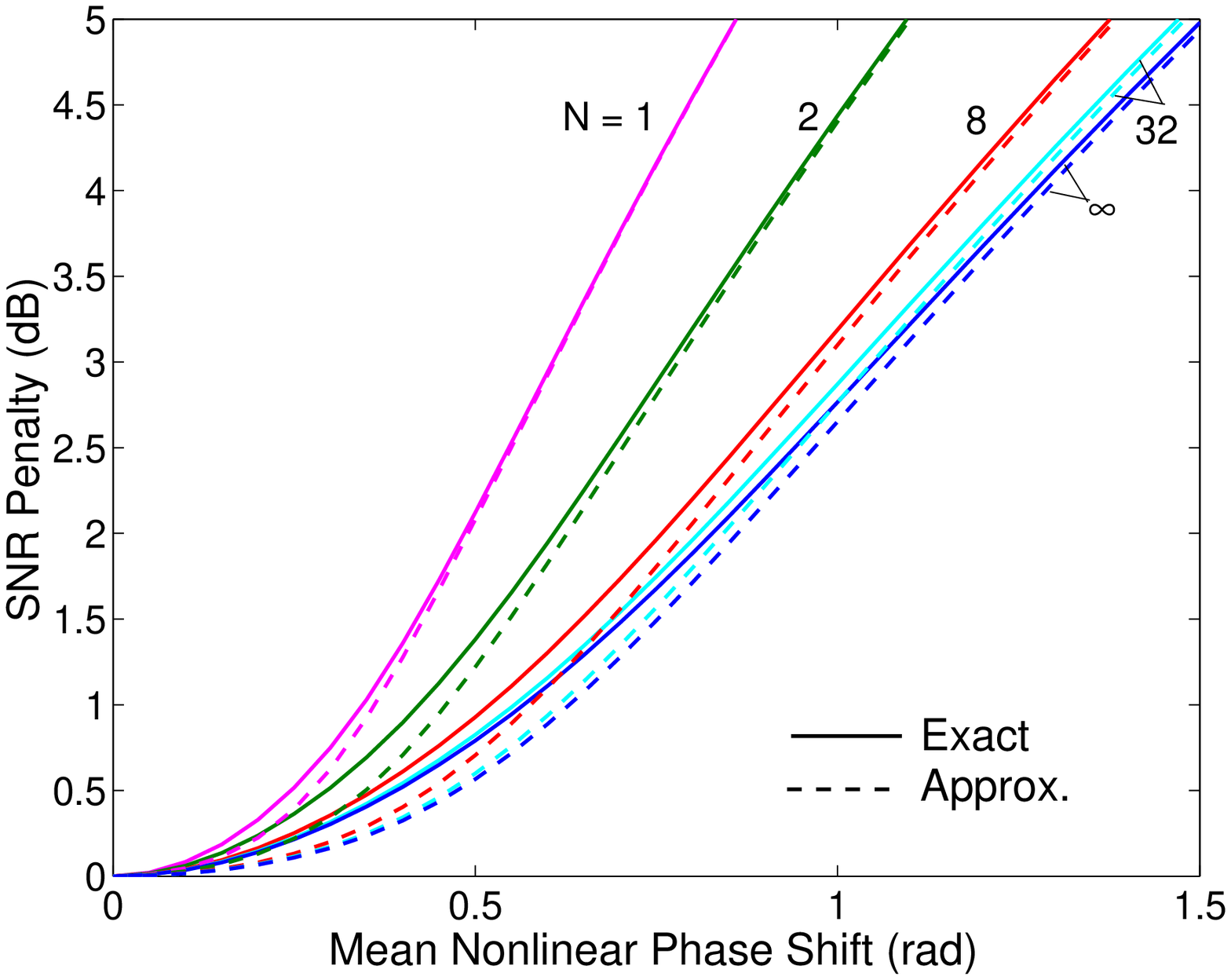}}
\caption{The SNR penalty vs. mean nonlinear phase shift  $<\!\!\Phi_\mathrm{NL}\!\!>$.}
\label{PenDPSKQ}
\end{figure}

For DPSK signals with quadratic phase noise, \Fig{BerDPSKQ} shows the exact error probability as a function of SNR $\rho_s$ for mean nonlinear phase shift of $<\!\!\Phi_\mathrm{NL}\!\!> = 0.5$ rad. 
\Fig{PenDPSKQ} shows the SNR penalty for an error probability of $10^{-9}$ as a function of mean nonlinear phase shift  $<\!\!\Phi_\mathrm{NL}\!\!>$.
The SNR penalty is defined as the additional required SNR to achieve the same error probability of $10^{-9}$.
Both Figs. \ref{BerDPSKQ} and \ref{PenDPSKQ} are calculated using \eqn{BERDPSK} and the independence approximation of \cite{ho0309b}.
The independence approximation of \cite{ho0309b} underestimates both the error probability and SNR penalty of a DPSK signal with quadratic phase noise of \eqn{phiNL}.
Both Figs. \ref{BerDPSKQ} and \ref{PenDPSKQ} also include the exact and approximated error probability for $N=\infty$ that are the distributed model from \cite{ho03exa} and \cite{ho0309a}, respectively.
The distributed model is applicable when the number of fiber spans is larger than 32.
In \fig{BerDPSKQ}, without quadratic phase noise of $<\!\!\Phi_\mathrm{NL}\!\!> = 0$, the error probability is $p_e = \exp(-\rho_s)/2$ \cite{proakis4}.
The required SNR for systems without nonlinear phase noise of $<\!\!\Phi_\mathrm{NL}\!\!> = 0$ is $\rho_s = 20$ (13 dB) for an error probability of $10^{-9}$.

From Figs. \ref{BerDPSKQ} and \ref{PenDPSKQ}, for the same mean nonlinear phase shift of $<\!\!\Phi_\mathrm{NL}\!\!>$, the SNR penalty is larger for smaller number of fiber spans.
When the mean nonlinear phase shift is $<\!\!\Phi_\mathrm{NL}\!\!> = 0.56$ rad, the SNR penalty is about 1 dB with large number ($N > 32$) of fiber spans but up to 3-dB SNR penalty for small number ($N = 1, 2$) of fiber spans.
For 1-dB SNR penalty, the mean nonlinear phase shift is also reduced from 0.56 to 0.35 rad with small number of fiber spans.

In \cite{gordon90}, the optimal operating point is defined when the variance of quadratic phase noise is approximately equal to the variance of the phase of Gaussian noise.
In \cite{ho0309a, ho0309b}, the optimal operating is calculated rigorously at the operation condition in which the increase of launched power does not improve the system performance.
The optimal operating point is reduced from 0.97 to 0.55 rad with the decrease of the number of fiber spans.

When the exact error probability is compared with the independence approximation of \cite{ho0309b}.
The independence approximation is closer to the exact error probability for small number of fiber spans.
In all cases, the independence assumption of \cite{ho0309b, ho0309a} underestimates the error probability of the system, contradicting to the conservative principle of system design.
The dependence between linear and nonlinear phase noise increases the SNR penalty up to 0.23 dB.

From the SNR penalty of \fig{PenDPSKQ}, if a prior penalty of about 0.23 dB is added into the system, the independence assumption of \cite{ho0309b} can be used to provide a conservative system design guideline.

\section{Conclusion}
\label{sec:end}

For a system with small number of fiber spans, the exact error probability of a DPSK signal with quadratic phase noise is derived analytically the first time when the dependence between linear and nonlinear phase noise is taking into account.
For the same mean nonlinear phase shift, the error probability increases for small number of fiber spans. 
The dependence between linear and nonlinear phase noises increases the error probability for DPSK signals.
Depending on the number of fiber spans, the SNR penalty increases by up to 0.23 dB due to the dependence between Gaussian noise and the quadratic phase noise. 

For the same mean nonlinear phase shifts and SNR, the error probability of the system increases with the decrease of the number of fiber spans.
As an example, the optimal operating point for system with large number ($N>32$) is a mean nonlinear phase shift of about 1 rad that is reduced to about 0.55 rad for system with small number of fiber spans ($N = 1, 2$). 
 

\end{document}